\documentstyle[aps,multicol,prl]{revtex}

\draft
\title{Kink Pairs Unbinding on Domain Walls and the Sequence of Phase
Transitions \\ in Fully Frustrated XY Models}
\author{S. E. Korshunov}
\address{L.D.Landau Institute for Theoretical Physics,
Kosygina 2, Moscow 117940, Russia}
\date{April 19, 2002}

\begin{document}
\maketitle

\begin{abstract}
The unbinding of kink pairs on domain walls in the fully frustrated $XY$
model (on square or triangular lattices) is shown to induce the vanishing
of phase coupling across the walls.
This forces the phase transition, associated with unbinding of vortex
pairs, to take place at a lower temperature than the other phase
transition, associated with proliferation of the Ising-type domain
walls. The results are applicable for a description of superconducting
junction arrays and wire networks in a perpendicular magnetic field,
as well as of planar antiferromagnets with a triangular lattice.

\end{abstract}

  \begin{multicols}{2}

A fully frustrated (FF) $XY$ model can be defined by the Hamiltonian
\begin{equation}
H=-J\sum_{(ij)}\cos(\varphi_{j}-\varphi_{i}-A_{ij})\;,     \label{1}
\end{equation}
where $J>0$ is the coupling constant, the fluctuating variables
$\varphi_{i}$ are defined on the sites $i$ of some regular
two-dimensional lattice, and the summation is performed over the pairs
of nearest neighbors $(ij)$ on this lattice. The non-fluctuating
(quenched) variables $A_{ij}\equiv -A_{ji}$  defined on lattice bonds
have to satisfy the constraint $\sum A_{ij}=\pi$
(where the summation is performed over the perimeter of a plaquette)
on all plaquettes of the lattice.

For two decades such models (on various lattices) have been
extensively studied in relation with experiments on Josephson junction
arrays \cite{ML}, in which $\varphi_{i}$ can be associated
with the phase of the superconducting order parameter
on the $i$-th superconducting grain, and $A_{ij}$ is related
to the vector potential of a perpendicular magnetic field,
whose magnitude corresponds to a half-integer number
of superconducting flux quanta per lattice plaquette.
A planar antiferromagnet with a triangular lattice also can be described
by the Hamiltonian (\ref{1}) (with $A_{ij}\equiv\pm\pi$).

The ground states of the FF $XY$ models on square \cite{Vil}
and triangular \cite{MS} lattices are characterized
by the \mbox{$U(1)\times Z_2$} degeneracy,
which suggests the possibility of two different phase transitions.
One of them (the Berezinskii-Kosterlitz-Thouless transition
\cite{Ber,KT,Kost}) can be associated with unbinding of vortex pairs
and the other with proliferation of the Ising type domain walls.

Teitel and Jayaprakash \cite{TJ} have proposed
that the temperature
$T_{\rm V}$ of the vortex pairs dissociation cannot be higher than
the temperature $T_{\rm DW}$ of the phase transition
associated with domain walls proliferation.
The arguments supporting this conjecture
have been put forward in Refs. \onlinecite{Hals,KU,K} and
are related to the presence on corners of domain walls of
fractional vortices, 
which are expected to screen the interaction of integer vortices
at $T>T_{\rm DW}$.

The application of the Hubbard-Stratanovich transformation \cite{HS} to
the FF $XY$ model on a square lattice allows one to reduce it \cite{CD} to
the system of two coupled unfrustrated $XY$ models,
which in the limit of strong coupling becomes equivalent
to the so-called $XY$-Ising model:
\begin{equation}
H_{}=-K\sum_{(ij)}(1+s_i s_j)
          \cos(\varphi_{i}-\varphi_{j})\;,                  \label{3}
\end{equation}
where $s_i=\pm 1$ is the auxiliary Ising type variable. In this model,
the coupling of the phase variables $\varphi_{i}$ across any domain
wall is completely absent. Although this property appears as a direct
consequence of taking (without any justification) the strong coupling
limit, and such a description fails to take into account
the existence of fractional vortices,
it has been suggested \cite{GKLN} that
the $XY$-Ising model {\em may} turn out to be a reasonable
approximation for investigation of the FF $XY$ model.

In the present Letter, we demonstrate that,
in the FF $XY$ model on square or triangular lattices,
the phase transition on a single domail wall, which takes place at $T_{\rm
K}<T_{\rm V}$ and consists in dissociation of pairs of logarithmically
interacting kinks \cite{LLKC}, induces for $T>T_{\rm K}$ the loss of
phase coupling across the wall.
This indeed makes the behavior of the FF $XY$ model analogous to that
of the $XY$-Ising model. We also show that the suppression of the phase
coupling between different Ising domains leads to $T_{\rm V}<T_{\rm
DW}$, at least when the phase transition associated with
domain wall proliferation is a continuous one.

The FF $XY$ model on square (or triangular) lattice being one of the
simplest examples of a system with non-perturbative coupling between
continuous and discrete degrees of freedom, the results are of interest
not only in relation to experimental realizations mentioned above,
but in a more general context of two-dimensional statistical mechanics.
In particular, we discuss in the conclusion their consequences for
the interplay between the roughening and the reconstruction transitions
\cite{dN,DdN}.
We do not consider here the FF $XY$ models on honeycomb \cite{K}
and dice \cite{dice} lattices, which are characterized by
much more developed discrete degeneracies.

In the ground state of the FF $XY$ model on a square lattice,
the gauge invariant phase differences
\makebox{$\theta_{ij}=\varphi_{j}-\varphi_{i}-A_{ij}$} on all bonds
are equal [when reduced to the interval $(-\pi,\pi)$] to $\pm\pi/4$
in such a way that summation of $\theta_{ij}$ over the perimeter of
each plaquette gives $\pi\sigma$, where $\sigma=\pm 1$ is called
chirality. The plaquettes with positive and negative chiralities
regularly alternate with each other, forming the checkerboard pattern
\cite{Vil}. The discrete twofold degeneracy of the ground state
corresponds to the change of the signs of all chiralities.

A domain wall can be defined as a topological excitation separating two
ground states which cannot be transformed into each other by a
continuous rotation. Schematically, it can be represented as a line on a
lattice, each link of which separates two plaquettes with the same
chirality (Fig. 1). A domain wall is characterized by a finite
energy per unit length;
therefore at low temperatures all domain walls which
appear as thermal fluctuations form closed loops.

If one considers an infinite straight domain wall and fixes the state
(the values of $\varphi_{i}$) on one side of the wall, the state on the
other side of the wall cannot be arbitrary and depends on the position
and on the orientation of the wall \cite{Hals,K}. If the wall is
displaced by one lattice constant, the values of $\varphi_{i}$ on the
other side of the wall are changed by $\pi$.

The presence of a kink (of the minimal height) on a domain wall [Fig.
1(a)] produces a mismatch of $\pi$ between the states which have to be
obtained when crossing the left and the right parts of the wall. This
discrepancy has to be taken care of by a fractional vortex with the
topological charge $\pm 1/2$ located on the kink. The energy of such
simple kink is therefore logarithmically divergent. The kinks of the
double (or, generally, even) height [Fig. 1(b)] do not introduce any
mismatch, and their energy is finite.


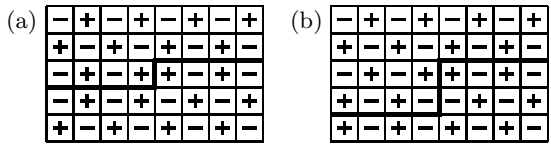
\begin{figure}[b]
\begin{center}

\newcounter{nnx}
\newcounter{nny}
\newcommand{\sta}[2]{\setcounter{nnx}{#1}\setcounter{nny}{#2}}
\newcommand{\ppp}{\multiput(\value{nnx},\value{nny})(-2.5,0){2}{\line(1,0){2.5}}
                  \multiput(\value{nnx},\value{nny})(0,-2.5){2}{\line(0,1){2.5}}\addtocounter{nnx}{10}}
\newcommand{\ppm}{\multiput(\value{nnx},\value{nny})(-2.5,0){2}{\line(1,0){2.5}}\addtocounter{nnx}{10}}

\setlength{\unitlength}{0.36mm}
\begin{picture}(215,56)(-20,-3)

\put(-15,42){\small (a)}
\put( 90,42){\small (b)}

\multiput(0,0)(0,10){6}{\line(1,0){80}}
\multiput(0,0)(10,0){9}{\line(0,1){50}}
\multiput(105,0)(0,10){6}{\line(1,0){80}}
\multiput(105,0)(10,0){9}{\line(0,1){50}}
{\thicklines
                                        \sta{5}{5}
\ppp \ppm \ppp \ppm \ppp \ppm \ppp \ppm \sta{5}{15}
\ppm \ppp \ppm \ppp \ppm \ppp \ppm \ppp \sta{5}{25}
\ppm \ppp \ppm \ppp \ppp \ppm \ppp \ppm \sta{5}{35}
\ppp \ppm \ppp \ppm \ppp \ppm \ppp \ppm \sta{5}{45}
\ppm \ppp \ppm \ppp \ppm \ppp \ppm \ppp

\put(00,20.3){\line(1,0){40.5}}
\put(00,19.7){\line(1,0){40.5}}
\put(39.5,29.7){\line(1,0){40.5}}
\put(39.5,30.3){\line(1,0){40.5}}
\put(39.7,19.7){\line(0,1){10.6}}
\put(40.3,19.7){\line(0,1){10.6}}

                                        \sta{110}{5}
\ppp \ppm \ppp \ppm \ppp \ppm \ppp \ppm \sta{110}{15}
\ppp \ppm \ppp \ppm \ppm \ppp \ppm \ppp \sta{110}{25}
\ppm \ppp \ppm \ppp \ppp \ppm \ppp \ppm \sta{110}{35}
\ppp \ppm \ppp \ppm \ppp \ppm \ppp \ppm \sta{110}{45}
\ppm \ppp \ppm \ppp \ppm \ppp \ppm \ppp

\put(105,10.3){\line(1,0){40.5}}
\put(105,09.7){\line(1,0){40.5}}
\put(144.5,29.7){\line(1,0){40.5}}
\put(144.5,30.3){\line(1,0){40.5}}
\put(144.7,09.5){\line(0,1){21.0}}
\put(145.3,09.5){\line(0,1){21.0}}}
\end{picture}
\end{center}
\caption[Fig. 1]{A domain wall with (a) a simple kink; (b) a double
kink. Pluses and minuses show the signs of chiralities.}
\end{figure}

Let us consider an infinite domain wall, introduced, for example, by
an appropriate choice of boundary conditions.
At low temperatures, it should contain a finite concentration of free
double kinks, but all simple kinks have to form neutral pairs.
Therefore, although the fluctuations of the domain wall diverge,
the symmetry with respect to its shift by one lattice constant is
broken.

With an increase of temperature, the phase transition in the
one-dimensional logarithmic gas of simple kinks will lead to
dissociation of neutral pairs and to the appearance of a finite
concentration of free simple kinks \cite{LLKC}. As follows from
the renormalization group analysis of Ref. \onlinecite{Bulg}, this
takes place when the prefactor of the logarithmic interaction
of simple kinks is equal to 2T, that is, at
\begin{equation}
T_{\rm K}=\frac{\pi}{4}\Gamma(T_{\rm K})\;.              \label{TK}
\end{equation}
Here $\Gamma(T)$ is the helicity modulus, the macroscopic parameter
describing the effective stiffness of the system with respect to
continuous twist of $\varphi_{}$.
At zero temperature $\Gamma(0)=J/\sqrt{2}$, whereas the unbinding of
integer vortices "in the bulk" takes place at \cite{NK}
\begin{equation}
T_{\rm V}=\frac{\pi}{2}\Gamma(T_{\rm V})\;,                 \label{UP}
\end{equation}
that is, above $T_{\rm K}$.
The numerical evidence for unbinding of kink pairs at $T_{\rm K}<T_{\rm V}$
has been obtained by Lee {\em et al.} \cite{LLKC}, who, however, mistook
this phase transition for the roughening transition of domain walls.

The phase transition associated with unbinding of pairs of simple kinks
leads to a restoration of the symmetry between the odd and the
even positions of the domain wall and also to the loss of the effective
phase stiffness across the wall. Any attempt to create a phase gradient
perpendicularly to the wall will be relaxed due to motion of free simple
kinks along the wall in different directions (in accordance with the sign
of the topological charge) under the action of Magnus force. The situation
is analogous to what happens in the bulk above $T_{\rm V}$,
when the presence of free vortices prevents the creation
of any stationary phase gradient (supercurrent).
For $T<T_{\rm K}$ all simple kinks are bound in neutral pairs and
their relative displacement requires a finite energy, which means
that the phase stiffness across the wall remains finite.

Analogously, a phase gradient parallel to the wall will not penetrate
on the other side of the wall.
Instead there will appear a difference in concentration of simple kinks
with positive and negative topological charges, which will compensate
for the difference in phase gradients on both sides of the wall.
Although creation of such difference in concentrations requires
some energy, this energy is proportional to the length of the
wall, whereas penetration of the phase gradient across the wall would
require the additional energy which is proportional to the total area
of the domain on the other side of the wall.
The same happens on grain boundaries in crystals, where the difference
in orientations is taken care of by a sequence of dislocations of the same
sign.

Note that both mechanisms work only at length scales which are
large in comparison with the inverse linear concentration of free
simple kinks. Nonetheless, their existence implies that at large
length scales the FF $XY$ model indeed can be approximated by the
$XY$-Ising model as proposed in Refs. \onlinecite{CD,GKLN}. Naturally,
at \makebox{$T<T_{\rm DW}$} such equivalence works only in a small
vicinity of $T_{\rm DW}$, in which the correlation length, defined in
terms of $\sigma$, is much larger than the typical distance between
free simple kinks on a  domain wall.

The same conclusions are valid in the case of the antiferromagnetic
$XY$ model on a triangular lattice. The ground state of this model
consists of three sublattices, the values of $\varphi_{i}$ in which
differ by $\pm 2\pi/3$, and also is characterized by
the $U(1)\times Z_2$ degeneracy \cite{MS}. If the ground state on one
side of a straight domain wall is fixed, the state on the other side of
the wall can be obtained by a permutation of values of $\varphi_{i}$ on
any two sublattices and subsequent rotation of all variables by
$\pi$ \cite{KU}.

The three available options correspond to three positions of the wall
and differ from each other by global rotation by $\pm 2\pi/3$.
Therefore, the simple kinks separating the straight parts of a
domain wall have to behave as fractional vortices with topological
charges $\pm 1/3$.
Accordingly, the phase transition associated with kink pairs unbinding
on an isolated infinite domain wall takes place at
$T_{\rm K}^{\triangle}
=({\pi}/{9})\Gamma(T_{\rm K}^{\triangle})$,
which is again below $T_{\rm V}$.
As in the case of a square lattice this phase transition leads
to the loss of phase coupling across a domain wall and makes
the behavior of the system analogous to that of the $XY$-Ising model.

The helicity modulus $\Gamma$ can be defined, in particular,
through the response of a system to application of the specially
chosen boundary conditions (see, for example, Ref. \cite{KNKB}).
The important property of the $XY$-Ising model is that, as soon as
there is at least one domain wall crossing the whole system,
the variables $\varphi_{}$ at opposite boundaries are completely
decoupled from each other, which means $\Gamma\equiv 0$.
In the thermodynamic limit this takes place
at any temperature higher than $T_{\rm DW}$.

At $T<T_{\rm DW}$ an externally imposed twist of $\varphi_{}$
in any typical configuration of $s_i$ can be carried
only by the largest (infinite) cluster formed by the sites with
the same sign of $s_i$ [note that the variable $s_i$ of the
$XY$-Ising model (\ref{3}) should be associated with the staggered
chirality of the FF $XY$ model and not with the chirality itself].
All other clusters have finite size
and therefore are insensitive to boundary conditions.

In two dimensions (in contrast to three), the point of the phase
transition of the Ising model coincides with the percolation transition
in the system of spin clusters \cite{CNPR}, so the density of the
infinite cluster decays on approaching $T_{\rm DW}$ from below as
$\rho(T)\propto\xi_p^{-\Delta d}$. Here $\Delta d=2-\overline{d}=5/96$
\cite{SV} is the deviation of the fractal dimension $\overline{d}$ of
the infinite cluster (at $T=T_{\rm DW}$) from its Euclidean dimension
and $\xi_p(T)\propto (T_{\rm DW}-T)^{-\nu}$ is the percolation length,
the temperature dependence of which in the Ising model is described by
the same exponent $\nu=1$ \cite{KSRC} as that of the correlation length
$\xi$.

Therefore the bare ({i.e.}, not reduced by the fluctuations of
$\varphi_{}$) helicity modulus $\Gamma_0(T)$ on approaching $T_{\rm
DW}$ has to decrease algebraically: $\Gamma_0(T)\propto(T_{\rm
DW}-T)^t$, at least as fast as $\rho(T)$ (actually much faster), which
can be shown with the help of the variational calculation. The vortex
pairs dissociation takes place as soon as the (renormalized) value of
$\Gamma$ is reduced to $(2/\pi)T$, that is, below $T_{\rm DW}$. For
$T_{\rm DW}\ll T_{\rm V}^{(0)}$ [where $T_{\rm
V}^{(0)}\approx (\pi/2)\Gamma (0)$ is the naive estimate for $T_{\rm
V}$], one therefore can expect
$T_{\rm DW}-T_{\rm V}
\propto[T_{\rm DW}/{T_{\rm V}^{(0)}}]^{1/t}\,.$
The two transitions can happen simultaneously only if they occur
as the first-order phase transition with larger than universal jump
in $\Gamma$.

According to our results the analogous behaviour can be expected in the
FF $XY$ models. However in that case the dependence of $T_{\rm
DW}-T_{\rm V}$ on $T_{\rm DW}$ will be more complicated, since the
effective reduction to the $XY$-Ising model is developed only at the
length scales which are large in comparison with typical distance
between free simple kinks. The idea that $\Gamma_0(T)$ is
strongly suppressed on approaching $T_{\rm DW}$ is supported by
a comparison of the results of numerical simulations of the same system
at zero and full frustration \cite{TJ}, which shows that in the latter
case the drop of $\Gamma$ with an increase of temperature
for the same size of the system is more sharp.

In conclusion, until
now the use of the $XY$-Ising model for the description of the
properties of the FF $XY$ model on a square lattice
\cite{GKLN,NGK} could be considered a rather arbitrary procedure.
Since application of the Hubbard-Stratanovich transformation
\cite{HS} to the FF $XY$ model on a triangular lattice is known
to produce a coarse-grained Hamiltonian \cite{CD}
with a wrong symmetry of the ground state
[$U(1)\times Z_3$ instead of $U(1)\times Z_2$],
one always could have doubts if the application of the same
transformation on square lattice does not lead to the loss of
some important properties of the original model.

An argument in favor of such an approach has been put forward by Knops
{\em et al.} \cite{KNKB}, who have shown with the help of the numerical
transfer matrix diagonalization that at $T=T_{\rm DW}$ the free energy
of the 19-vertex version of the FF $XY$ model with increasing the
system size becomes insensitive to the boundary conditions inducing the
twist of $\varphi$. The authors of Ref. \cite{KNKB} have interpreted
this as evidence for irrelevance at criticality of the coupling of
$\varphi$ across a domain wall. The same observation  could be
alternatively interpreted as evidence for $T_{\rm V}<T_{\rm DW}$.

In the present Letter we have demonstrated that in the FF $XY$ models
on square and triangular lattices
the loss of the phase coupling across domain walls
is achieved already at temperatures below $T_{\rm V}$
due to the presence on domain walls of free simple kinks.
We also have shown that the dissociation of vortex pairs has to take
place at $T_{\rm V}<T_{\rm DW}$,
because, on approaching $T_{\rm DW}$ from below, the part of the system
which reacts to external twist becomes more and more dilute \cite{DU}.
This makes the scenario of a single phase transition with a novel
critical behavior \cite{KU,LKG} impossible, at least for
$T>T_{\rm K}$.

These conclusions are not dependent on the particular form of the
interactions in the system (as soon as the degeneracy of the ground state
remains the same), and are applicable, for example, also when the
interaction of further neighbors is taken into account \cite{FCKF}.
They are in agreement with the results of the most recent numerical
simulations of the FF $XY$ models on square \cite{Ols,LL1} and
triangular \cite{LL2} lattices, as well as of the equivalent
(half-integer) Coulomb gas \cite{Lee} and of the SOS-Ising model
\cite{LLK} which is (partially) dual to the $XY$-Ising model.

Recently Lee {\em et al.} \cite{LLK} have demonstrated that the generalized
$XY$-Ising model is dual to the SOS-Ising model introduced by den Nijs
\cite{dN} for the coarse-grained description of the interplay between
roughening and missing row reconstruction on a surface of a crystal
with simple cubic lattice. The proposed phase diagram of this model can
be found in Fig. 3 of Ref. \onlinecite{DdN}. For the case of $\Delta=0$
(which corresponds to the $XY$-Ising model with a complete absence of phase
coupling across domain walls) it contains two regions, one (at
$R<0$) with separated and the other (at $R>0$) with coinciding phase
transitions. Our analysis, as well as the results of the numerical
simulations of Ref. \onlinecite{LLK}, implies that the two transitions
should be separated for both signs of $R$.

It seems worthwhile to mention that the conclusion on a larger than
universal value of $\Gamma(T_{\rm V})$ in the FF $XY$ models
\cite{LL1,LL2,Lee} has been obtained with the help of the
Weber-Minnhagen (WM) scaling analysis \cite{WM}, which is based on the
same renormalization group equations \cite{Kost} as the universal
prediction (\ref{UP}) and, accordingly, does not even allow for a
possibility of a nonuniversal value of $\Gamma(T_{\rm V})$. The
results of Refs. \cite{LL1,LL2,Lee} therefore should be interpreted as
evidence for deviation from the WM scaling. The reasons for such
a deviation can be easily understood. It follows from our analysis
that, even when the presence of vortex pairs (the only factor taken into
account in the framework of the WM analysis) is neglected, $\Gamma$ in
the vicinity of $T_{\rm DW}$ has to be strongly scale dependent,
because the effects related with suppression of the effective stiffness
across domain walls develop only with the increase of scale. Thus it is
important not to confuse the two mechanisms for suppression of
$\Gamma$. The method for plotting the data, which allows for checking in
what interval of length scales the WM scaling really holds,
has been recently suggested in Ref. \onlinecite{FCKF}.


The author is grateful to L. N. Shchur for useful comments.
This work has been supported by the Program "Quantum Macrophysics"
of the Russian Academy of Sciences,
by the Program "Scientific Schools of the Russian Federation"
(grant No. 00-15-96747),
by the Swiss National Science Foundation and
by the Netherlands Organization for Scientific Research (NWO)
in the framework of Russian-Dutch Cooperation Program.

  \end{multicols}
\end{document}